\documentclass[lettersize,journal]{IEEEtran}
\usepackage{amsmath,amsfonts}
\usepackage{algorithmic}
\usepackage{algorithm}
\usepackage{array}
\usepackage[labelfont=normalfont, labelsep=period]{caption}
\usepackage{textcomp}
\usepackage{stfloats}
\usepackage{url}
\usepackage{verbatim}
\usepackage{graphicx}
\usepackage{cite}
\usepackage{xcolor}
\usepackage{graphicx}
\usepackage{booktabs}
\usepackage{multirow}
\usepackage{subcaption}
\usepackage[colorlinks,linkcolor=blue,citecolor=blue]{hyperref}
\begin{document}

\title{Enhancing Discrete Particle Swarm Optimization for Hypergraph-Modeled Influence Maximization}

\markboth{Journal of \LaTeX\ Class Files,~Vol.~14, No.~8, August~2015}
{Shell \MakeLowercase{\textit{et al.}}: A Sample Article Using IEEEtran.cls for IEEE Journals}


\author{Qianshi Wang, Xilong Qu, Wenbin Pei, Nan Li, Qiang Zhang,~\IEEEmembership{Senior Member,~IEEE} 
\thanks{  
This work was supported in part by the National Key Research and Development Program of China under grant 2021ZD0112400, the National Natural Science Foundation of China under grant 62206041, the 111 Project under grant D23006, China University Industry-University-Research Innovation Fund under grants 2022IT174, Natural Science Foundation of Liaoning Province under grant 2023-BSBA-030, and an Open Fund of National Engineering Laboratory for Big Data System Computing Technology under grant SZU-BDSC-OF2024-09. (Corresponding author: Wenbin Pei.)  
}
 \thanks{Xilong Qu, Wenbin Pei, and Qiang Zhang are with the School of Computer Science and Technology, Dalian University of Technology, Dalian 116024, China; Key Laboratory of Social Computing and Cognitive Intelligence (Dalian University of Technology), Ministry of Education, Dalian 116024, China (e-mail: quxilong@mail.dlut.edu.cn; peiwenbin@dlut.edu.cn;  zhangq@dlut.edu.cn).

 Qianshi Wang is with the International School of Information Science \& Engineering, Dalian University of Technology, Dalian 116024, China (e-mail: W\_Qianshi@outlook.com). 

Nan Li is with the School of Artificial Intelligence, Shanxi University, Taiyuan 237016, China. Nan Li is also with the Institute of Big Data Science and
Industry, and College of Software, Northeastern University, Shenyang 110819, China. (email: lnnner@163.com)
}
}

\maketitle

\begin{abstract}
Influence maximization (IM) is a fundamental problem in complex network analysis, with a wide range of real-world applications. To date, existing approaches to influential node identification in IM have predominantly relied on standard graphs, failing to capture higher-order intrinsic interactions embedded in many real-world systems. Hypergraphs can be employed to better capture higher-order interactions. However, using hypergraphs may lead to an excessively large search space and increased complexity in modeling cascading dynamics, making it challenging to accurately identify influential nodes. Therefore, in this study, we propose a new hypergraph-modeled IM method, based on the Discrete Particle Swarm Optimization algorithm and the threshold model. In the proposed method, a particle (i.e., a candidate solution) represents the selection information of seed nodes, and the fitness function is designed to accurately and efficiently evaluate the influence of seed nodes via a two-layer local influence approximation. We also propose a degree-based initialization strategy to improve the quality of initial solutions and develop rules for updating particles' velocity and position, incorporated with a local search to drive particles toward better solutions. Experimental results demonstrate that the proposed method outperforms baseline methods on both synthetic and real-world hypergraphs. In addition, ablation studies validate the effectiveness of both the local search and the initialization strategies.
\end{abstract}


\section{Introduction}
\IEEEPARstart{M}{any}real-world complex systems often consist of an extremely large number of nodes. For example, in social networks \cite{peng2011technology}, millions of users are interconnected through billions of social ties, serving as critical infrastructures for interpersonal communication \cite{eginli2018interpersonal} and large-scale information dissemination \cite{bartoloni2024twenty}. 
However, due to budgetary and resource constraints, organizations tend to select only a small number of initial users, who are expected to initiate large-scale information cascades in order to maximize marketing influence \cite{yanchenko2024influence}. 
This can be formally defined as an influence maximization (IM) problem \cite{kempe2003maximizing}, which aims to identify a subset of nodes (referred to as ``seed nodes") with the highest potential to propagate influence throughout the network. The IM problem has been proven to be NP-hard \cite{Zhang2024influence} and has a broad range of real-world applications, including viral marketing \cite{song2014influence,banerjee2020survey}, epidemic control \cite{cheng2020outbreak,aral2018social}, and information dissemination \cite{zhang2014recent,zhang2022influence,tang2018online}.


Most existing studies on the IM problem are based primarily on graphs \cite{li2023survey,kumar2022influence,heidari2015smg}, failing to effectively capture higher-order relationships among nodes \cite{Zhang2024influence}. For example, in real-world social networks, individuals frequently engage in group activities or belong to communities, giving rise to higher-order interactions that are fundamentally group-oriented rather than strictly pairwise. Ordinary graphs, which represent relationships as binary connections between pairs of nodes, are hard to accurately model such group interactions \cite{li2018influence}. Differently, hypergraphs \cite{battiston2020networks} provide a more expressive modeling paradigm, where a single hyperedge can connect any number of nodes simultaneously, enabling a more accurate representation of higher-order dependencies among nodes. Therefore, the use of hypergraphs allows for a more precise modeling of information dissemination processes embedded in the complex group interactions observed in real-world systems \cite{de2020social}. 


However, due to the structural complexity of hypergraphs, the IM problem based on hypergraphs faces two fundamental challenges in the accurate identification of influential nodes\cite{zhu2018social}\cite{antelmi2021social}. The first is the vast search space because a network typically consists of an extremely large number of nodes and edges\cite{li2023influence}. Moreover, hyperedges that connect multiple nodes simultaneously further compound this complexity by capturing group-based interactions that exponentially increase the number of relationships\cite{wang2024hedv}.

Particle swarm optimization (PSO) \cite{kennedy1995particle}, recognized as one of the most prominent and well-established algorithms in evolutionary computation, has demonstrated remarkable effectiveness in addressing complex combinatorial optimization problems. 
Unlike the genetic algorithm, which relies on complex genetic operators like crossover and mutation, PSO achieves faster convergence with fewer parameters and a simpler optimization framework. Therefore, this study proposes a PSO-based framework tailored for the hypergraph-modeled IM problem, which aims to efficiently and accurately identify influential nodes while fully leveraging the higher-order interactions inherent in hypergraphs.
The main contributions of this study are summarized as follows:

\begin{enumerate}

\item[1)] 
This is the first study to apply PSO to the IM problem modeled by hypergraphs. To guide the search efficiently, we design a two-layer local influence approximation to estimate the spread of a given seed set under the threshold model.



\item[2)] The new velocity update rule in PSO is enhanced to better capture the structural characteristics of hypergraphs. Furthermore, a local search strategy is integrated. This enables the proposed method to escape local optima, contributing to more effective and robust identification of influential nodes than existing methods.



\item[3)] 
Extensive experiments conducted on both synthetic hypergraphs and real-world hypergraphs demonstrate that the proposed method can effectively identify influential node sets and consistently outperforms the state-of-the-art baseline methods, demonstrating its superior effectiveness and scalability in capturing higher-order interactions in complex networks.


\end{enumerate}

\section{Background}
\subsection{The Problem Formulation of IM}
IM is a fundamental optimization problem in complex network analysis, with important applications in areas such as epidemic control \cite{cheng2020outbreak}, and product marketing \cite{huang2019community,li2020substantial}. The IM problem aims to identify a subset \( S \subseteq V \) of \( K \) nodes that maximizes the expected influence spread \( \sigma(S) \) under a given propagation model. This objective can be formulated as:
\begin{equation}
\begin{aligned}
\arg\max_{S \subseteq V} \ \sigma(S) \\
\text{s.t.} \quad |S| = K
\end{aligned}
\end{equation}
\subsection{Related Works}

Early studies on IM mainly employed centrality-based heuristics, including betweenness centrality \cite{brandes2001faster}, closeness centrality \cite{okamoto2008ranking}, and PageRank \cite{brin1998anatomy}, as primary measures for quantifying node importance. These studies are computationally efficient, while they often exhibit limited precision because they fail to capture the dynamic nature of information propagation. To improve accuracy, more advanced strategies have been proposed. For example, iterative methods such as HADP \cite{xie2023efficient} enhance propagation performance by selecting seed nodes with low influence overlap. Greedy algorithms, e.g., CELF \cite{kempe2003maximizing} and CELF++ \cite{goyal2011celf++}, achieve high precision by directly modeling the propagation process. However, their high computational cost limits their scalability to large-scale networks.

Evolutionary algorithms (EAs) have been increasingly adopted to address the IM problem due to their strong global search capabilities. Representative methods include discrete particle swarm optimization (DPSO) \cite{gong2016influence}, the discrete bat algorithm (DBA) \cite{tang2018maximizing}, the hybrid frog leaping algorithm \cite{tang2020discrete}, and adaptive-based evolutionary models (ABEM) \cite{li2023abem}. These methods strike a good trade-off between efficiency and performance. However, most of these methods rely on traditional graph models, which are insufficient for capturing the higher-order interactions commonly observed in real-world networks.

With the rising interest in higher-order network modeling, the IM problem in hypergraphs has garnered increasing research attention. Unlike graphs, hypergraphs allow a hyperedge to connect any number of nodes simultaneously, enabling more accurate modeling of group interactions. 
Several works \cite{puzis2013betweenness,xie2023efficient} focused on redefining node centrality measures and designing heuristic algorithms tailored to hypergraph structures. For example, Xie et al. \cite{xie2023efficient} proposed a degree-based adaptive algorithm that achieves low computational complexity while maintaining competitive performance on both synthetic and real-world hypergraphs. Puzis et al. \cite{puzis2013betweenness} extended the concept of betweenness centrality to hypergraphs; however, this method suffers from a high computational overhead, limiting its scalability to large networks. 
Qu et al. \cite{qu2024influence} introduced genetic algorithms to address the IM problem on hypergraphs, demonstrating their superior performance over the degree-based or centrality-based methods. 
However, most existing methods still struggle to effectively capture the complex propagation dynamics and the strong interdependence between nodes and hyperedges inherent in hypergraph models\cite{ma2022hyper}. 

\section{Preliminary}
In this section, we introduce the fundamental concepts of hypergraphs, followed by an introduction to the propagation model and PSO.

\subsection{Hypergraphs}
\begin{figure}[htbp]
    \centering
    \includegraphics[width=\columnwidth]{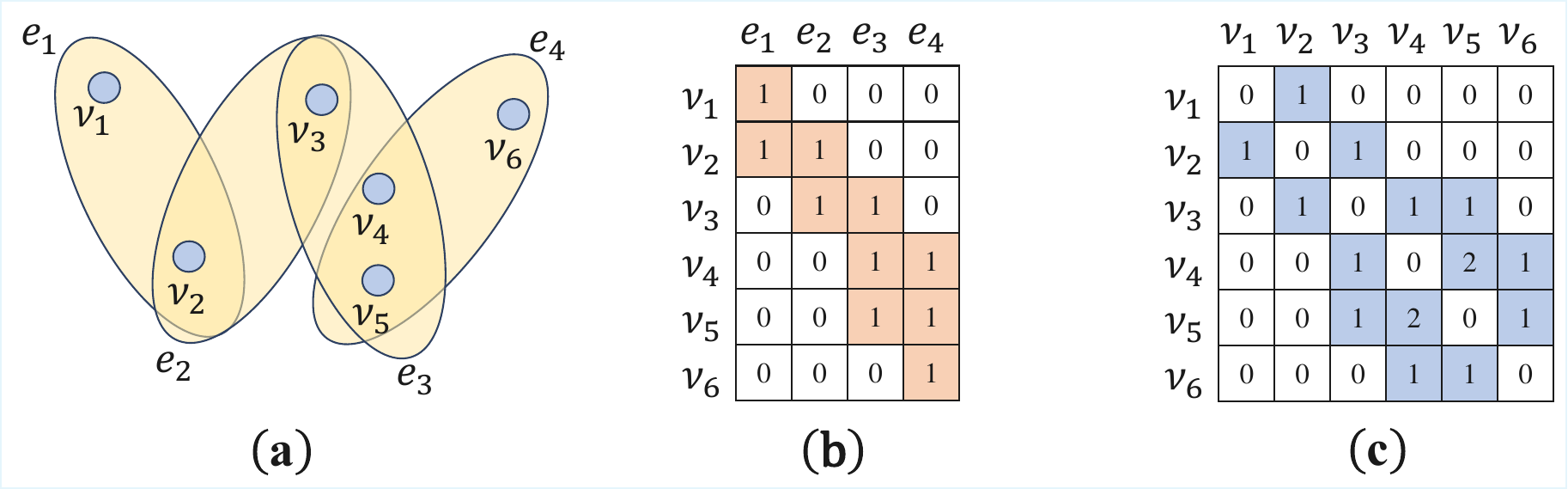} 
    \caption{An example hypergraph. (a) Hypergraph H; (b) the incidence matrix of H; (c) the adjacency matrix of H.}
    \label{fig:hypergraphimage}
\end{figure}

A hypergraph $H = (V, E)$ consists of a set of vertices $V = \{v_1, v_2, \dots, v_n\}$ and a set of hyperedges $E = \{e_1, e_2, \dots, e_m\}$, where each hyperedge $e_j \subseteq V$ captures a high-order relationship among the involved vertices. A hypergraph can be represented by an $n \times m$ incidence matrix $C$. An element $c_{ij}$ in $C$ is defined as:

\begin{equation}
c_{ij} =
\begin{cases}
1, & \text{if } v_i \in e_j, \\
0, & \text{otherwise}.
\end{cases}
\end{equation}
Based on the incidence matrix, the adjacency matrix \( A \in \mathbb{R}^{n \times n} \) of the hypergraph can be computed as:
\begin{equation}
  A = C C^T - D,  
\end{equation}
where \( D \) is a diagonal matrix with \( D_{ii} \) denoting the degree of node \( v_i \), i.e., the number of hyperedges that contain \( v_i \). \( A_{ij} \) indicates the number of hyperedges containing both \( v_i \) and \( v_j \). An example hypergraph \( H \) is illustrated in Fig.~\ref {fig:hypergraphimage}.

The degree of a vertex \( v_i \), denoted as \( \deg(v_i) \), represents the number of distinct neighbors of \( v_i \) in the hypergraph, defined as:
\begin{equation}
\deg(v_i) = \sum_{j=1}^{n} \mathcal{A}_{ij},
\end{equation}
where \( \mathcal{A} \in \mathbb{R}^{n \times n} \) is derived from the incidence matrix \( C \) as follows:
\begin{equation}
\mathcal{A}_{ij} =
\begin{cases}
1, & \text{if } (C C^\top)_{ij} > 0, \\
0, & \text{otherwise}.
\end{cases}
\end{equation}

Here, \( \mathcal{A}_{ij} = 1 \) indicates that vertices \( v_i \) and \( v_j \) share at least one common hyperedge. The hyperdegree of node $v_j$ is defined as the number of hyperedges that contain $v_j$: 
\begin{equation}
d_H(v_j) = \sum_{i=1}^{m} C_{ij}.
\end{equation}

Based on the above definitions, the degree and hyperdegree of the nodes in Fig.~\ref{fig:hypergraphimage} can be calculated. For example, the degree of node \( v_4 \) equals 3, and the hyperdegree of node \( v_4 \) equals 2.

\subsection{Propagation Models}

\begin{figure}[htbp]
    \centering
    \includegraphics[width=\columnwidth]{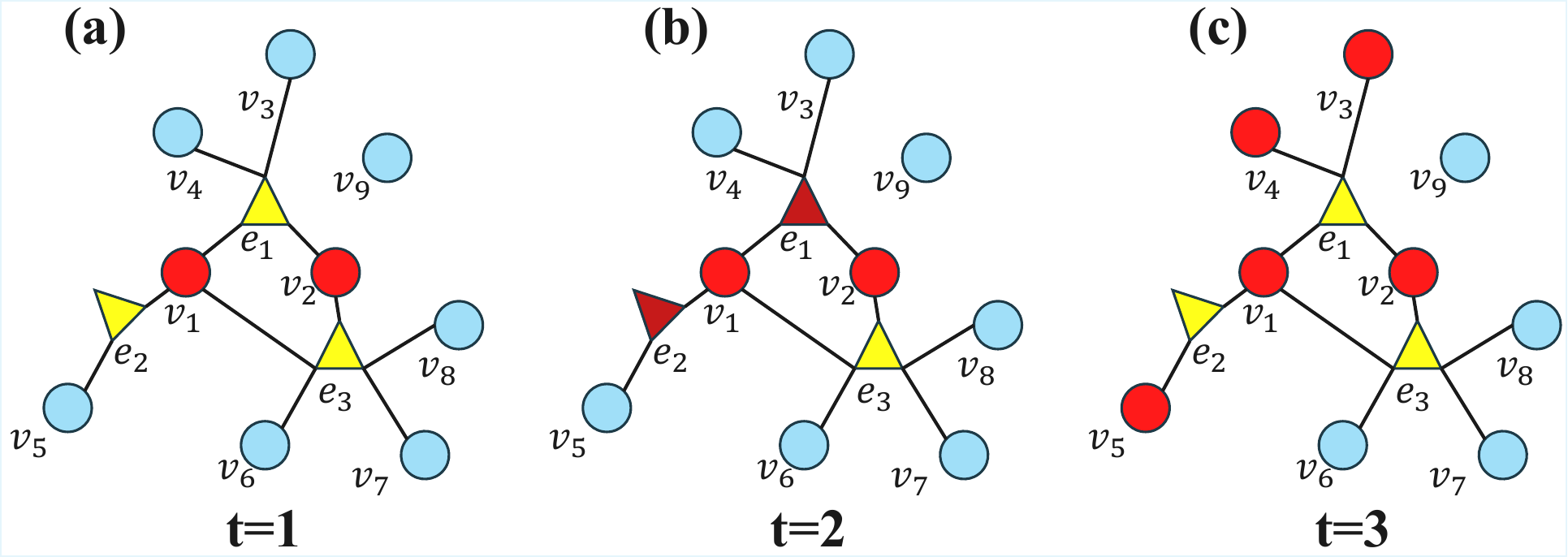}
    \caption{An example of propagation in a single threshold model. Circles denote nodes, and triangles denote hyperedges. Activated nodes and hyperedges are marked in red and dark red, respectively. }
    \label{fig:threshimage}
\end{figure}

The threshold model is one of the most classical and widely studied diffusion models~\cite{singh2013threshold}. Despite its simplicity, it effectively captures various real-world propagation phenomena, such as opinion formation and behavioral adoption~\cite{wei2013influence}. In this study, we adopt the threshold model~\cite{Zhang2024influence} as the underlying propagation mechanism to quantify the influence spread initiated by a given seed set. Given an initial seed set $S$, the influence propagation process in the hypergraph is defined as follows:


\begin{itemize}
  \item \textbf{Step 1:} Initially, all hyperedges are inactive. Nodes in the seed set \( S \) are marked as active, while all other nodes remain inactive.
  
  \item \textbf{Step 2:} Iterate over all inactive hyperedges. For each hyperedge \( e_j \), if the ratio of active nodes within \( e_j \) exceeds or equals a given threshold \(p\), then \( e_j \) is activated and all nodes contained within  \( e_j \) are simultaneously activated.

  \item \textbf{Step 3:} Repeat Step 2 iteratively until no new hyperedges can be activated.
\end{itemize}
Fig.~\ref{fig:threshimage} depicts the propagation process of the threshold model with a threshold value \( p = 0.5 \).

\subsection{Particle Swarm Optimization}

PSO is a population-based metaheuristic, inspired by the social behavior of bird flocks. 
PSO was originally developed for continuous optimization and was later extended to discrete domains in \cite{kennedy1997discrete} through the introduction of Binary PSO (BPSO). Numerous variants have been introduced to enhance performance across diverse application areas, among which the improved version in \cite{shi1998modified} is widely adopted. 

In PSO, each particle in a swarm maintains a position vector representing a candidate solution and a velocity vector that governs its movement through the solution space.
Given the seed set size  \( k \) and the swarm size \( n \), for particle \( i \) (\( i = 1, 2, \ldots, n \)), let \(\mathbf{X}_i = (x_{i1}, x_{i2}, \ldots, x_{ik})\) and \(\mathbf{V}_i = (v_{i1}, v_{i2}, \ldots, v_{ik})\) denote the position and velocity vectors of particle $i$, respectively. The velocity and position are updated according to the following equations:
\begin{equation}
    \mathbf{V}_i \leftarrow \omega \mathbf{V}_i + c_1 r_1 (\mathbf{Pbest}_i - \mathbf{X}_i) + c_2 r_2 (\mathbf{Gbest} - \mathbf{X}_i)
\end{equation}

\begin{equation}
    \mathbf{X}_i \leftarrow \mathbf{X}_i + \mathbf{V}_i
\end{equation}
Here \( \omega \) is the inertia weight, \(\mathbf{Pbest}_i = (pbest_{i1}, pbest_{i2}, ..., \\pbest_{ik})\) is the personal best position of particle \( i \), and \(\mathbf{Gbest} = (gbest_1, gbest_2, ..., gbest_k)\) is the global best position in a swarm. The parameters \( c_1 \) and \( c_2 \) are learning factors, while \( r_1 \) and \(r_2 \) are randomly generated values within \([0,1]\).  

\section{The Proposed Method}
\begin{figure*}[htbp]  
    \centering
    \includegraphics[width=\textwidth]{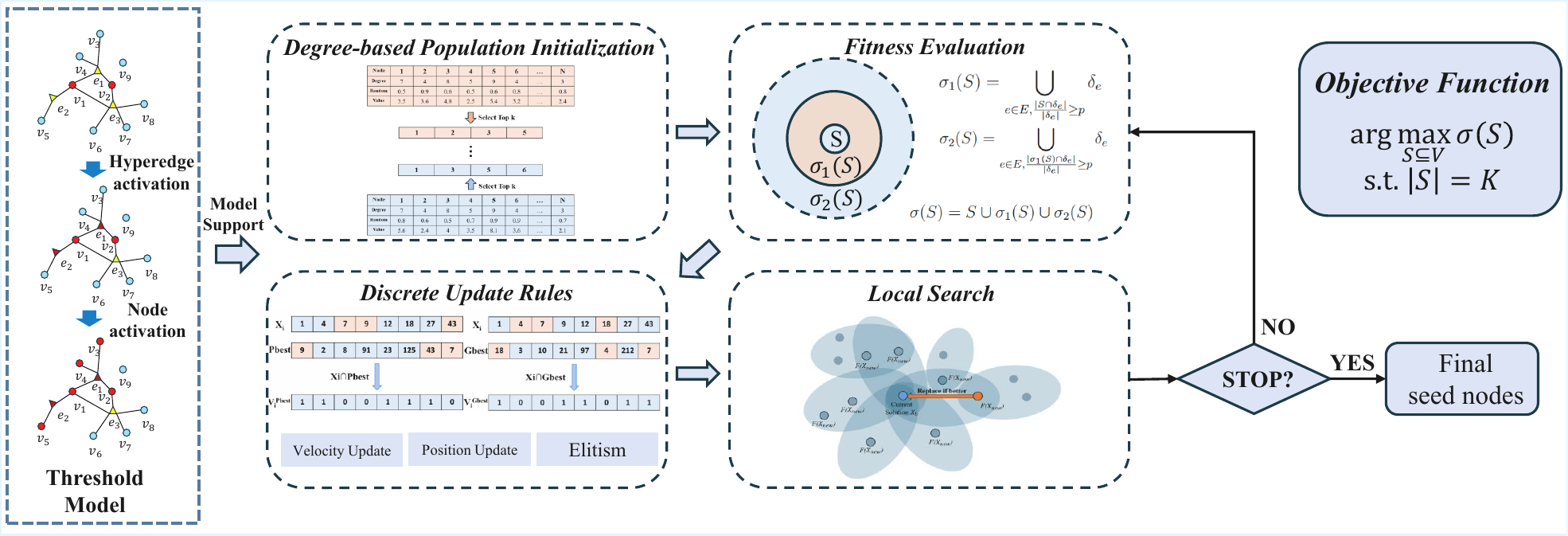} 
    \caption{The Flowchart of the HDPSO Method.}
    \label{fig:frame}
\end{figure*}

In this section, we propose a new method, named Hypergraph-based Discrete Particle Swarm Optimization (HDPSO), for identifying the most influential seed nodes in an IM problem. 
The flowchart of HDPSO is illustrated in Fig.~\ref{fig:frame}. This method begins with a degree-based population initialization method. Then, the spread influence of individuals (i.e., seed sets) is preliminarily evaluated to guide the search. The particles are then updated through two procedures. The first procedure uses standard update rules in PSO. The second procedure adopts a local search strategy, aiming to enhance convergence while maintaining diversity. 
The evaluation and population-updating cycle iterates continuously until the predefined termination criteria are satisfied. 


The following subsections provide a detailed description of the individual representation, initialization strategy, update rules, and the local search procedure.

\subsection{Individual Encoding}



In the HDPSO method, a particle is composed of $k$ seed nodes. Both the position and velocity vectors of particles are redefined in a discrete manner to adapt the method for solving the IM problem in hypergraphs. The specific encoding scheme is described as follows.

In HDPSO, the position of particle \( i \) is represented as a \( k \)-dimensional integer vector \( X_i = (x_{i1}, x_{i2}, \ldots, x_{ik}) \) (where \( i = 1, 2, \ldots, n \)), corresponding to a candidate seed set for influence maximization within the hypergraph. Each element \( x_{ij} \) (where \( x_{ij} \leq |V| \) and \( j = 1, 2, \ldots, k \)) denotes the index of a selected node. For example, if the seed set size is \( k = 5 \) and the position vector of particle \( i \) is \( X_i = (1, 3, 6, 7, 9) \), it indicates that nodes 1, 3, 6, 7, and 9 are selected as the seed set in the hypergraph.

The velocity of particle \( i \) is defined as \( V_i = (v_{i1}, v_{i2}, \ldots, v_{ik}) \), a binary update indicator rather than a continuous displacement. During each iteration, particles update their velocities based on both the global best solution and their historical best position, thereby determining the search direction toward a better solution. Each component \( v_{ij} \in \{0, 1\} \) indicates whether the corresponding node in the position vector should be updated (\( v_{ij} = 1 \)) or retained (\( v_{ij} = 0 \)). This representation naturally matches the combinatorial nature of seed selection, where a solution is a subset of nodes rather than a point in continuous space.

\subsection{The Degree-based Population Initialization Method}

To accelerate the convergence of the proposed HDPSO method, we design a degree-based initialization method for generating the position vectors of particles. 

First, each node is assigned a degree-based score based on its degree. Note that although selecting the nodes with the highest degrees can improve the quality of initial solutions, this strategy risks premature convergence if particles start from similar positions. To address this issue, we introduce a random factor \( \xi_i \) into the degree-based score. 
Specifically, for each node \( v_i \), we compute its score by the following equation:
\begin{equation}
S(v_i) = \deg(v_i) \times \xi_i,\quad \xi_i \sim \mathcal{U}(0.5, 1.0),
\end{equation}
where \( \deg(v_i) \) denotes the degree of node \( v_i \), and \( \xi_i \) is a uniformly distributed random number in the interval \( (0.5, 1.0] \). 

All nodes are then ranked in descending order based on \( S(v_i) \), after which the top \( k \) nodes are selected to form the initial seed set for each particle.




\subsection{The Fitness Function}

Evaluating the influence of a given seed set in the threshold model typically requires simulating the cascading process across the entire network, which requires considerable computational cost. 
Suggested by the previous study \cite{pei2014searching}, a node’s global influence is predominantly determined by its two-hop neighborhood. Therefore, in a hypergraph $H=(V,E)$, to approximate the influence spread of the seed set $S$, we evaluate the influence of $S$ using the structural information of the two-hop neighborhood. This avoids repeated global propagation, substantially reducing computational overhead compared to Monte Carlo simulation. 
More details are introduced as follows:


First, we define a two-layer local influence approximation to estimate the spread of a given seed set $S$ under the threshold model. Let $\delta_e$ denote the set of nodes associated with hyperedge $e$, and $p \in (0,1]$ be the uniform activation threshold. The first layer of activated nodes, denoted as $\sigma_1(S)$, is defined as:
\begin{equation}
\sigma_1(S) = \bigcup_{e \in E, \frac{|S \cap \delta_e|}{|\delta_e|} \geq p} \delta_e
\end{equation}

This layer includes all the nodes belonging to hyperedges where the fraction of seed nodes is greater than or equal to the threshold $p$, triggering the activation of the entire hyperedge.

Subsequently, the second layer of activated nodes, denoted as $\sigma_2(S)$, is computed by applying the same rule to the newly activated nodes from $\sigma_1(S)$:
\begin{equation}
\sigma_2(S) = \bigcup_{e \in E, \frac{|\sigma_1(S) \cap \delta_e|}{|\delta_e|} \geq p} \delta_e
\end{equation}

The fitness function $f(S)$ is then defined as the total number of nodes activated after the second layer of propagation:
\begin{equation}
\sigma(S) = S \cup \sigma_1(S) \cup \sigma_2(S)
\end{equation}

\begin{equation}
f(S) = |\sigma(S)|
\end{equation}

The HDPSO method aims to identify a seed set \( S \) of size \( k \) for maximizing the function \( f(S) \). This approximation balances estimation accuracy and computational efficiency, enabling efficient evaluation of individuals during the evolutionary process.

\subsection{Update Rules}



In PSO, particles' velocity vectors are crucial in guiding solutions toward promising regions in the search space. In HDPSO, the velocity update rule is reformulated in a discrete form as follows:
\begin{equation}
    \mathbf{V}_i \gets \mathbf{D} \left( \omega \mathbf{V}_i + c_1 r_1 (\mathbf{Pbest}_i \cap \mathbf{X}_i) + c_2 r_2 (\mathbf{Gbest} \cap \mathbf{X}_i) \right)
    \label{con:PSO}
\end{equation}

\begin{figure}
    \centering
    \includegraphics[width=\columnwidth]{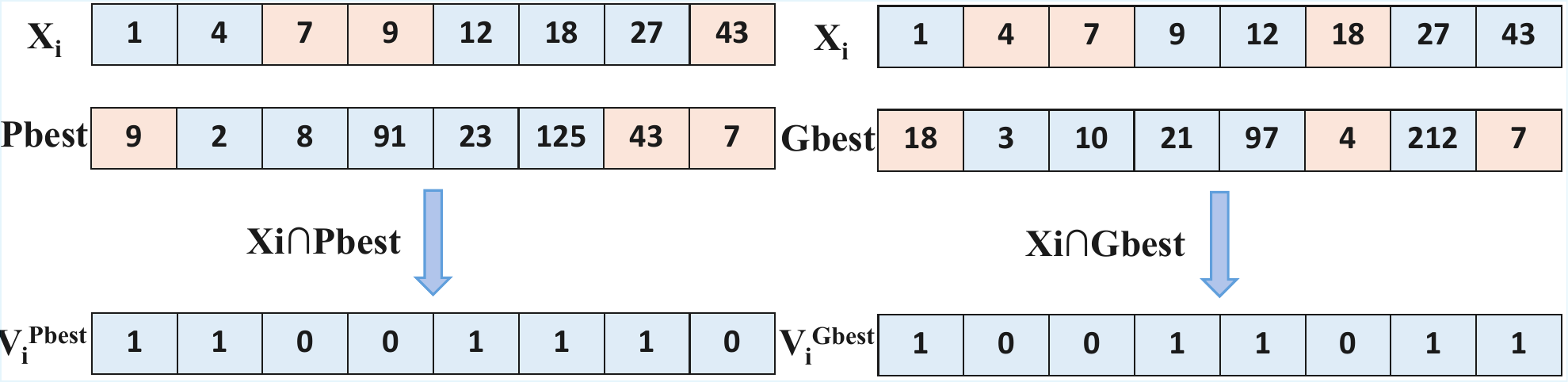}
    \caption{Example of velocity update in the discrete PSO. Given \(c_1 r_1 = 0.6\) and \(c_2 r_2 = 1.1\), the velocity vector \(\mathbf{V}_i\) is computed as $\mathbf{V}_i = \mathbf{D}(0.6 \times (1,1,0,0,1,1,1,0) + 1.1 \times (1,0,0,1,1,0,1,1)) = \mathbf{D}(1.7,0.6,0,1.1,1.7,1.7,0.6,1.7) = (1,0,0,1,1,1,0,1). $ }
    \label{fig:updateimage}
\end{figure}

Here, the operator \(\cap\) denotes a similarity-based intersection operation. The intersection operation is used to measure similarity between solutions, allowing particles to inherit common elements from Pbest and Gbest. An illustrative example is shown in Fig.~\ref{fig:updateimage}. In this example, given the seed set size of \(k=8\), the position vector of particle \(i\) is given as \(X_i = (1,4,7,9,12,18,27,43)\). The personal best position is \(\mathbf{Pbest}_i = (9,2,8,91,23,125,43,7)\), and the global best is \(\mathbf{Gbest} = (18,3,10,21,97,4,212,7)\).

To calculate the intersection, we identify the common elements between \(X_i\) and \(\mathbf{Pbest}_i\), which are nodes 7, 9, and 43. The corresponding entries in the velocity vector \(V_i^{\mathbf{Pbest}}\) are set to 0, indicating these components should be retained, while other entries are set to 1. Thus,
\[
V_i^{\mathbf{Pbest}} = (1,1,0,0,1,1,1,0).
\]
Similarly, the global best vector \(V_i^{\mathbf{Gbest}} = (1,0,0,1,1,0,1,1)\) is obtained.

The function \(\mathbf{D}(\cdot)\) acts element-wise over the resulting weighted sum, defined as:
\[
\mathbf{D}(X_i) = \left(d_1(x_{i1}), d_2(x_{i2}), \dots, d_k(x_{ik})\right),
\]
where
\begin{equation}
d_j(x_{ij}) =
\begin{cases}
0, & x_{ij} < \tau, \\
1, & x_{ij} \geq \tau,
\end{cases}
\end{equation}
Here, $\tau \in [1,2]$ is a threshold parameter controlling whether the combined guidance from personal and global best positions triggers replacement.

Once the new velocity vector is obtained, the position of the particle is updated as:
\begin{equation}
\mathbf{X}_i' \leftarrow \mathbf{X}_i \circ \mathbf{V}_i
\end{equation}
Let the updated position vector be \(\mathbf{X}_i' = (x'_{i1}, x'_{i2}, \dots, x'_{ik})\). Afterward, for each element:
\begin{equation}
x'_{ij} =
\begin{cases}
x_{ij}, & \text{if } v_{ij} = 0; \\
\text{Replace}(x_{ij}, V), & \text{if } v_{ij} = 1.
\end{cases}
\end{equation}

Here, \(V\) represents the set of all nodes in the hypergraph \(H\), and the function \texttt{Replace}(\(x_{ij}, V\)) replaces \(x_{ij}\) with a randomly selected node from \(V\), ensuring uniqueness within the new vector.

At each iteration, the velocity vector incorporates personal and global experiences to guide particle movement. The updated velocity is then used to selectively alter components of the position vector, helping particles efficiently explore the solution space.

\subsection{The Local Search Strategy}




To accelerate the search process, we introduce a hypergraph-based local search strategy. The local search refines solutions by exploring the neighborhood structure of hypergraphs, helping PSO escape local optima that it might otherwise converge to prematurely. Specifically, the local search strategy performs a greedy search over neighboring nodes, replacing a selected node with a neighboring candidate if the substitution improves the particle's fitness.  
This strategy is applied to a subset of particles, effectively improving the overall convergence of a swarm. The detailed procedure is presented in \textbf{Algorithm~\ref{alg:local_search}}. In line 6 of \textbf{Algorithm~\ref{alg:local_search}}, $N_{x_{bi}}$ is the set of the neighboring nodes of $x_{bi}$. The \textit{Update} function is responsible for replacing $x_{bi}$ with a \( \textit{neighbor} \) in $X_b$. 

The proposed HDPSO method integrates population-based PSO and a local search strategy to enhance both exploration and exploitation. While PSO enables the swarm to identify multiple promising regions through learning and leader guidance, the local search operator refines solutions within these regions by leveraging neighborhood information. By combining these two approaches, the HDPSO method effectively balances global exploration with local refinement, speeding up convergence and improving solution quality in the IM problem in hypergraphs.

\begin{algorithm}
    \caption{Local\_Search($\mathbf{X_a}$)}
    \label{alg:local_search}
    \begin{algorithmic}[1]
        \STATE \textbf{Input:} Particle $\mathbf{X_a}$
        \STATE $\mathbf{X_b} \gets \mathbf{X_a}$;    
        \STATE Best\_fitness $\gets$ F($\mathbf{X_b}$);    ~~~\# $F(\cdot)$ represents the above fitness function that approximates the influence spread of a particle.
        \FOR{each element $x_{bi} \in \mathbf{X_b}$}
            \IF{random() $< 0.2$} 
                \FOR{$neighbor \in N_{x_{bi} }$}   
                    \STATE $\mathbf{X_{\text{new}}} \gets \text{Update}(\mathbf{x_{bi}}, \mathbf{neighbor})$;
                    \STATE New\_fitness $\gets$ F($\mathbf{X_{new}}$);
                    \IF{New\_fitness $>$ Best\_fitness}
                        \STATE $\mathbf{X_b} \gets \mathbf{X_{new}}$;
                        \STATE Best\_fitness $\gets$ New\_fitness;
                    \ENDIF
                \ENDFOR
            \ENDIF
        \ENDFOR
        \STATE \textbf{Output:} Particle $\mathbf{X_b}$.
    \end{algorithmic}
\end{algorithm}



\section{Experiment Designs}
\begin{table}[h]
    \caption{Statistical Characteristic Settings of Synthetic Hypergraphs}
    \label{tab:synthetic-hyper}
    \centering
    \begin{tabular}{ccccc}
        \hline
        Network & $N$   & $M$            & Feature & Threshold Value \\
        \hline
        ER      & 2000  & 1000, 2000     & $\langle k\rangle = 3, 4$ & 0.5 \\
        SF      & 2000  & 500, 1000      & $\lambda = -1.5, -2$ & 0.5 \\
        K-UF    & 2000  & 1000, 2000     & $m = 3, 4$ & 0.5 \\
        \hline
    \end{tabular}
\end{table}
\subsection{Hypergraphs}
To comprehensively evaluate the performance of the proposed method, we conducted experiments on both synthetic and real-world hypergraphs. For the synthetic hypergraphs, we generated Erd\H{o}s--R\'enyi (ER), scale-free (SF), and $k$-uniform (K-UF) hypergraphs using the configuration model in~\cite{xie2023efficient}, with detailed parameters listed in Table~\ref{tab:synthetic-hyper}. All experiments were conducted on the largest connected component of each hypergraph to eliminate the effects of isolated nodes.
To further validate the applicability of HDPSO in real-world scenarios, we selected four representative hypergraphs: Algebra (cat-edge-algebra-questions), Geometry (cat-edge-geometry-questions), Music-Blue (cat-edge-music-blues-reviews), and IJO1366. These datasets cover diverse application domains. The Algebra and Geometry hypergraphs represent collaborative interactions in online QA (Question and Answer) communities, while Music-Blue reflects user–product review behaviors in an e-commerce setting. The IJO1366 hypergraph models metabolic reactions in biological systems. The detailed parameters of real-world hypergraphs are represented in Table~\ref{tab:realworld}.



\begin{table*}  
    \centering
    \caption{Topological properties of the four real-world hypergraphs used in our experiments}
    \label{tab:realworld}
    \setlength{\tabcolsep}{15pt} 
    \begin{tabular}{lrrrrrrrrrr}
        \hline
        Hypergraph & $n$  & $m$   & $\langle\deg\rangle$ & $\langle d^H\rangle$ & $\langle d^E\rangle$ & $c$   & $\langle d\rangle$ & $\xi$ & $\rho$  & $t_{\gamma}$ \\
        \hline
        Algebra     &  423  & 1268  &  78.90  & 19.53  &  6.52  & 0.79  & 1.95  & 5     & 0.19   & 0.85 \\
        Geometry    &  580  & 1193  & 164.79  & 21.53  & 10.47  & 0.82  & 1.75  & 4     & 0.28   & 0.70 \\
        Music-Blue   & 1106  &  694  & 167.87  &  9.49  & 15.13  & 0.62  & 1.99  & 8     & 0.15   & 0.60 \\
        IJO1366     & 1805  & 2546  &  16.91  &  5.55  &  3.94  & 0.58  & 2.62  & 7     & 0.009  & 0.80 \\
        \hline
    \end{tabular}

     \begin{flushleft}    
\footnotesize     $n$ and $m$ denote the numbers of nodes and hyperedges, respectively; 
    $\langle\deg\rangle$ is the average node degree; 
    $\langle d^H\rangle$ is the average node hyperdegree; 
    $\langle d^E\rangle$ is the average hyperedge size; 
    $c$ is the clustering coefficient; 
    $\langle d\rangle$ is the average shortest-path length; 
    $\xi$ is the network diameter; 
    $\rho$ is the edge density of the corresponding ordinary graph; 
    $t_{\gamma}$ is the threshold value.
\end{flushleft}

\end{table*}

\subsection{Baseline Methods}
To evaluate the effectiveness of HDPSO, we compare it against seven representative baseline methods, introduced as follows:

\begin{enumerate}
    \item[1)] \textbf{GA} \cite{qu2024influence}: A standard GA is used to evolve a seed set of size \(K\);
    \item[2)] \textbf{High Hyper-Degree (HHD)} \cite{xie2023efficient}: A heuristic that iteratively selects the node with the highest hyper-degree until seed nodes are identified.
    \item[3)] \textbf{Random (RD)} \cite{Zhang2024influence}: It uniformly samples distinct nodes from the hypergraph as seeds.
    \item[4)] \textbf{Neighbor Priority (NP)} \cite{Zhang2024influence}: This is a centrality-based method that ranks nodes according to their number of neighbors, selecting the top as seeds.
    \item[5)] \textbf{PageRank} \cite{Zhang2024influence}: This is an adaptation of the classic PageRank centrality for hypergraphs. It selects the nodes with the highest scores as seeds.
    \item[6)] \textbf{HCI1} \cite{Zhang2024influence}: This is a first-order Hypergraph Collective Influence measure, which ranks nodes based on the number of subcritical paths of length one emanating from them.
    \item [7)] \textbf{HCI2} \cite{Zhang2024influence}: This is a second-order Hypergraph Collective Influence measure, which ranks nodes based on the number of subcritical paths of length two emanating from them.
\end{enumerate}
Note that each stochastic algorithm (including HDPSO, GA, and RD) was executed 30 times, and the results were averaged.
\subsection{Parameter Settings}

The threshold parameter is crucial for shaping the propagation dynamics in hypergraphs. Setting the parameter too high limits activation to a small set of nodes, which severely curtails influence spread and results in uniformly poor performance. Conversely, a low threshold leads to near-complete propagation, which diminishes the distinction between methods. To ensure a meaningful comparison, we fix the threshold at $0.5$ for the synthetic hypergraphs, allowing a balanced and controllable propagation process. For the real-world hypergraphs, where topological and behavioral characteristics vary considerably, we empirically select dataset-specific thresholds to better distinguish between different methods. The detailed threshold settings are summarized in Table~\ref{tab:realworld}.

All experiments were conducted on a PC equipped with a 5.6 GHz Intel i7 CPU. The proposed method was implemented in Python using the DEAP framework. To enhance the search efficiency and mitigate premature convergence, the local search probability was set to \( p_l = 0.1 \), meaning that local search in each generation targets 10\% of the swarm. To achieve a balance between exploration and exploitation, the threshold parameter $\tau$ is set to 1.5 to avoid overly frequent or overly conservative replacements. This design improves the method's exploration capabilities while maintaining a relatively low computational overhead. 
Parameter sensitivity experiments were conducted to identify the optimal parameter settings, as presented in Table~\ref{tab:pso_params}.


\begin{table}[htbp]
    \centering
    \caption{Parameter settings of HDPSO}
    \label{tab:pso_params}
    \begin{tabular}{lcl}
        \hline
        \textbf{Meaning} & \textbf{Parameter} & \textbf{Value} \\
        \hline
        Generations                   & \texttt{maxgen}   & 50   \\
        Population size                & \texttt{pop}      & 256  \\
        Cognitive component weight     & \texttt{c1}       & 1.2  \\
        Social component weight        & \texttt{c2}       & 1.2  \\
        Inertia weight                 & $\omega$          & 0.7  \\
        Local search probability       & $p_l$             & 0.1  \\
        Replacement threshold          & $\tau$            & 1.5  \\
        \hline
    \end{tabular}
\end{table}
\section{Results and Analysis}
\subsection{Experiments on the Synthetic Hypergraphs}
\begin{figure*}[htbp]  
    \centering
    \includegraphics[width=\textwidth]{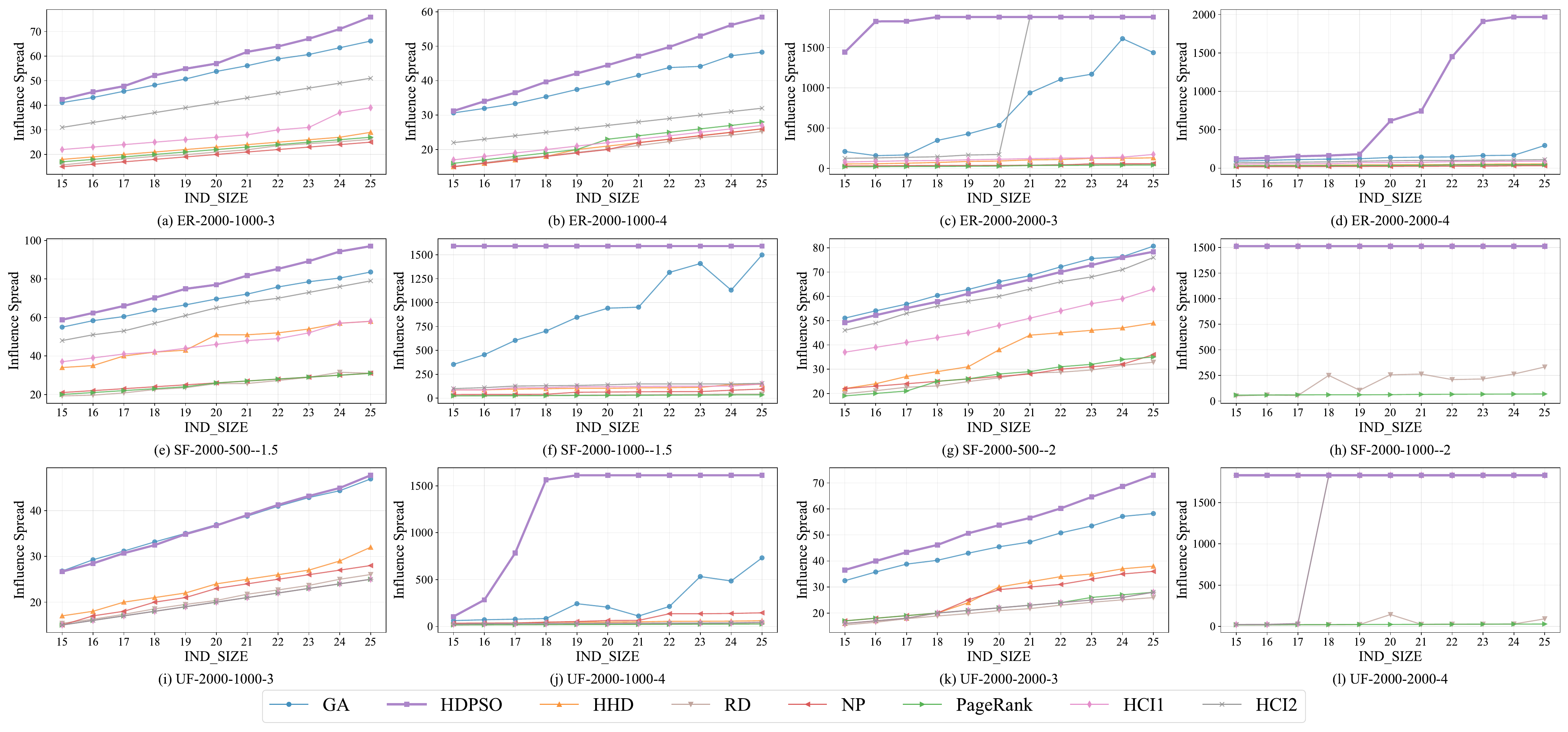} 
    \caption{The performance of HDPSO and the baseline methods on the synthetic hypergraphs. The horizontal axis represents the number of seed nodes, and the vertical axis represents the spread influence (the number of activated nodes).}
    \label{fig:synthetic}
\end{figure*}

Fig.~\ref{fig:synthetic} illustrates the results of the proposed HDPSO method compared with the baseline methods on the ER, SF, and K-UF hypergraphs. It is indicated that HDPSO demonstrates a significant performance advantage in almost all cases. 
As the number of seed nodes increases and the optimization landscape becomes more complex, the advantage of HDPSO becomes more pronounced, owing to its strong exploration capability in identifying high-quality solutions. On the SF-2000-500–2 hypergraph, HDPSO underperforms the standard GA, primarily because the hypergraph contains relatively few hyperedges, leading to lower structural complexity. 

\subsection{Experiments on the Real-world Hypergraphs}

\begin{figure}
    \centering
    \includegraphics[width=\columnwidth]{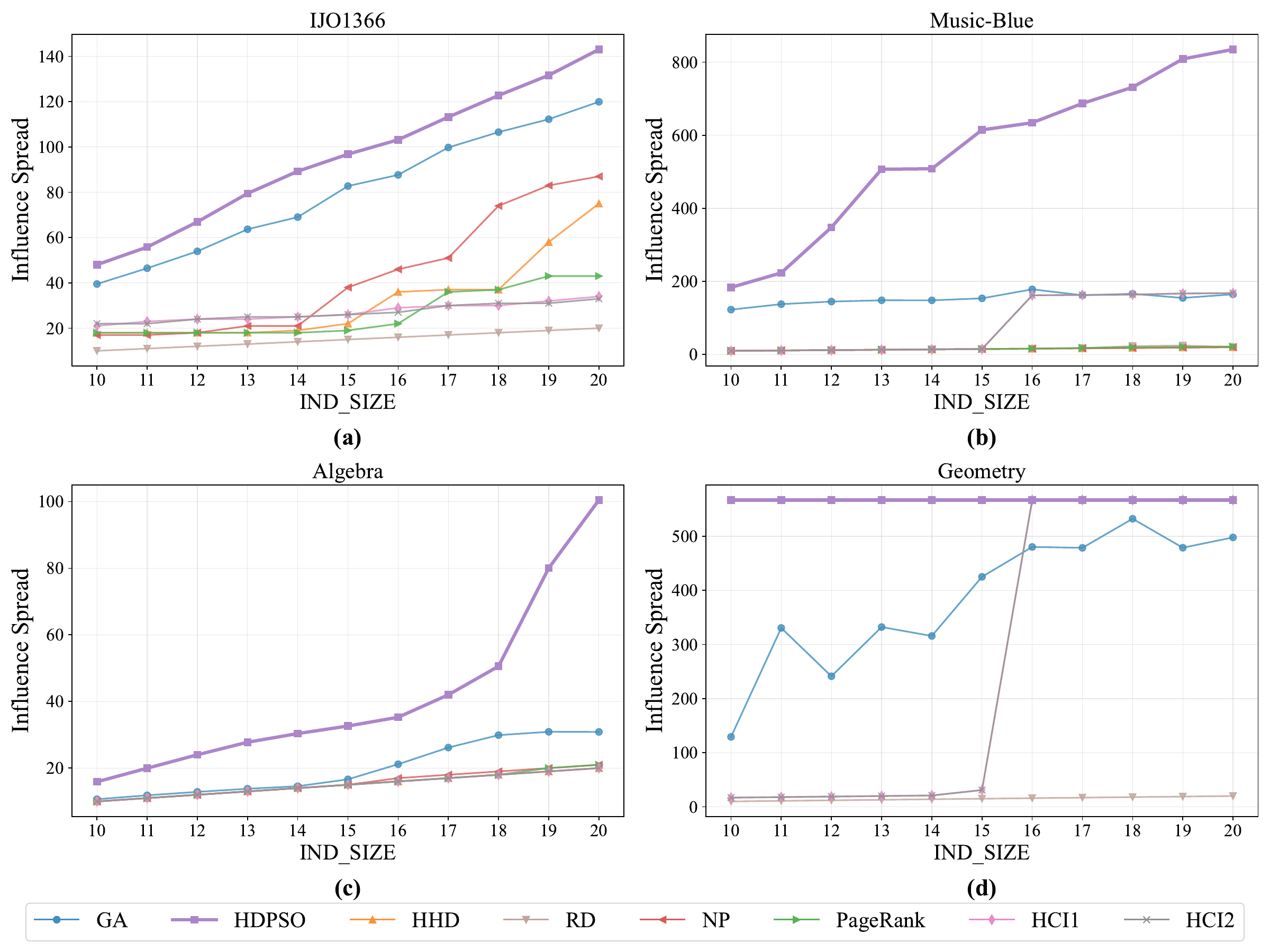}
    \caption{The performance of HDPSO and the baseline methods on the real-world hypergraphs. The horizontal axis represents the number of seed nodes, and the vertical axis represents the spread influence (the number of activated nodes).}
    \label{fig:synthetic-real}
\end{figure}

Fig.~\ref{fig:synthetic-real} illustrates the results of HDPSO and the baseline methods on the real-world hypergraphs. As shown in Fig.~\ref{fig:synthetic-real}, HDPSO consistently outperforms the baseline methods across all the datasets and varying seed set sizes. As the number of seed nodes increases, the superiority of HDPSO becomes more evident, highlighting the effectiveness of its global exploration capability and adaptive mechanisms in larger search spaces. Moreover, as a representative evolutionary algorithm, the GA exhibits relatively strong performance on the \textit{IJO1366} and \textit{Geometry} datasets; however, its influence spread remains inferior to that of HDPSO. On the structurally more complex \textit{Algebra} hypergraph, most methods, except HDPSO, exhibit limited propagation capability, highlighting the superior search ability of HDPSO. Notably, on the \textit{Geometry} dataset, the high density of node connections facilitates influence propagation. HDPSO achieves nearly complete propagation, selecting only 10 seed nodes. This indicates that the network structure has a significant impact on influence spread.

\subsection{Ablation Experiments}

\begin{figure*}[htbp]
    \centering
    \includegraphics[width=\textwidth]{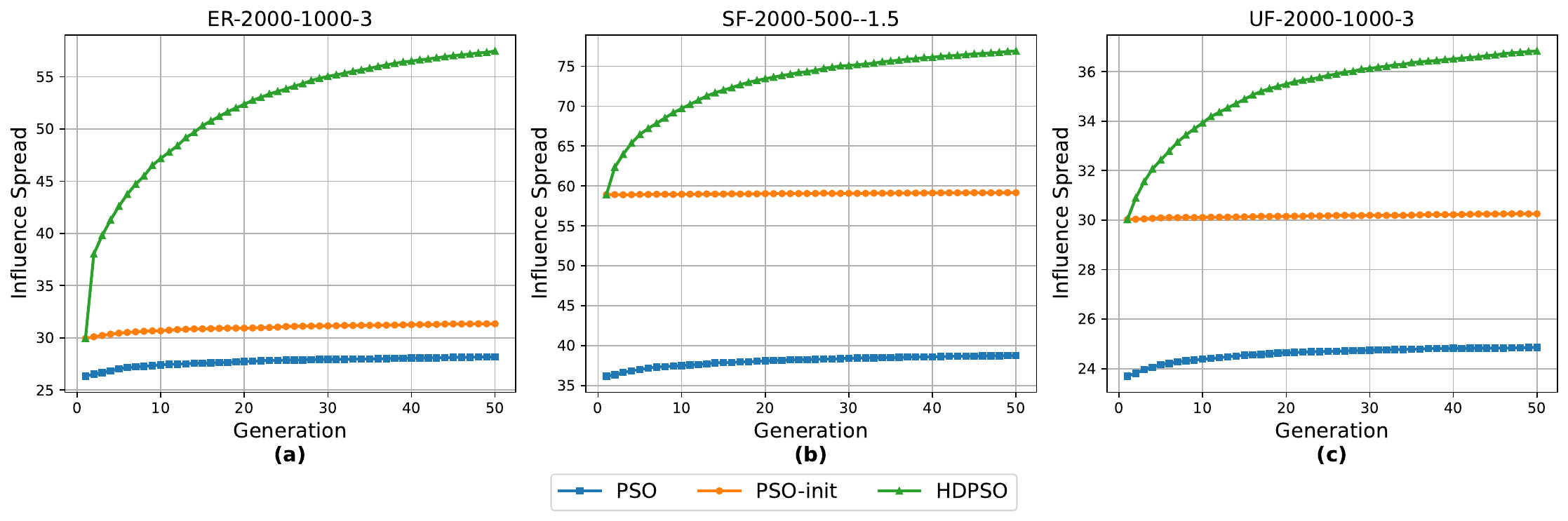}
    \caption{Illustration of the changes of propagation values with generations for PSO, PSO-init, and HDPSO methods. (a–c) represent the changes of fitness values on ER hypergraph, SF hypergraph, and K-UF hypergraph, respectively.}
    \label{fig:ablation}
\end{figure*}

To validate the effectiveness of the proposed components, we compare HDPSO with the following variants:

\begin{itemize}
    \item The standard PSO;
    \item PSO-init: a variant using only the proposed degree-based initialization strategy;
\end{itemize}
Table~\ref{tab:ablation} summarizes the hypergraph parameters in the ablation experiments. Fig.~\ref{fig:ablation} illustrates the changes of propagation values with generations for PSO, PSO-init, and HDPSO methods. As illustrated in Fig.~\ref{fig:ablation}, both PSO-init and HDPSO incorporate a degree-based influence strategy during initialization, resulting in higher propagation values than the standard PSO algorithm in the early iterations. This indicates that the designed initialization enables the proposed method to evolve high-quality solutions with fewer iterations. Furthermore, an analysis of the evolutionary processes of PSO-init and HDPSO reveals that the local search operator accelerates convergence, guiding the population toward better solutions. By the 50th generation, HDPSO demonstrates superior propagation performance compared to PSO-init, which itself outperforms the standard PSO. These results demonstrate that both the initialization and local search operators significantly improve solution quality for HDPSO.

\begin{table}[htbp]
    \centering
    \setlength{\tabcolsep}{3pt}  
    \caption{Statistical settings of hypergraphs used in ablation experiments}
    \label{tab:ablation}
    \begin{tabular}{ccccc}
        \hline
        Network & $N$ & $M$ & Feature & Threshold Value \\
        \hline
        ER     & 2000 & 1000 & Average Hyperdegree $\langle k\rangle = 3$ & 0.5 \\
        SF     & 2000 & 500  & Power-law exponents $\lambda = -1.5$        & 0.5 \\
        K-UF   & 2000 & 1000 & Cardinality of hyperedges $m = 3$          & 0.5 \\
        \hline
    \end{tabular}
\end{table}

\subsection{Statistical Significance Test Results}
We conducted Wilcoxon rank-sum tests with a significance level of 0.05 to evaluate the statistical significance of performance differences between the proposed HDPSO method and the baseline methods. 
Table~\ref{tab:synthetic} reports the detailed experimental results on the synthetic hypergraphs, and Table~\ref{tab:real} presents those on the real-world hypergraphs.

\begin{table*}[!t]
\centering
\caption{Results of HDPSO and Baseline Methods on All the Synthetic Hypergraphs}
\label{tab:synthetic}
\renewcommand{\arraystretch}{1.8}
\setlength{\tabcolsep}{4pt}
\resizebox{\textwidth}{!}{%
\begin{tabular}{|l|cccc|cccc|cccc|}
\hline
\multirow{2}{*}{Algorithm}
& \multicolumn{4}{c|}{ER}
& \multicolumn{4}{c|}{SF}
& \multicolumn{4}{c|}{UF} \\
\cline{2-13}
& ER-2000-1000-3 & ER-2000-1000-4 & ER-2000-2000-3 & ER-2000-2000-4
& SF-2000-500--1.5 & SF-2000-500--2 & SF-2000-1000--1.5 & SF-2000-1000--2
& UF-2000-1000-3 & UF-2000-1000-4 & UF-2000-2000-3 & UF-2000-2000-4 \\
\hline

GA
& 53.439$\pm$7.986~$\mathbf{+}$
& 39.361$\pm$5.847~$\mathbf{+}$
& 737.039$\pm$508.162~$\mathbf{+}$
& 145.403$\pm$52.690~$\mathbf{+}$
& 69.485$\pm$9.079~$\mathbf{+}$
& 65.845$\pm$9.297~$\mathbf{=}$
& 927.715$\pm$364.903~$\mathbf{+}$
& 1513.009$\pm$0.029~$\mathbf{=}$
& 36.936$\pm$6.199~$\mathbf{=}$
& 256.467$\pm$215.923~$\mathbf{+}$
& 45.697$\pm$8.179~$\mathbf{+}$
& 1831.000$\pm$0.000~$\mathbf{=}$ \\

PSO
& 29.812$\pm$4.094~$\mathbf{+}$
& 25.839$\pm$3.674~$\mathbf{+}$
& 97.279$\pm$19.139~$\mathbf{+}$
& 58.897$\pm$8.477~$\mathbf{+}$
& 41.882$\pm$4.982~$\mathbf{+}$
& 41.376$\pm$5.358~$\mathbf{+}$
& 97.412$\pm$29.889~$\mathbf{+}$
& 1517.785$\pm$1.473~$\mathbf{=}$
& 25.603$\pm$3.935~$\mathbf{+}$
& 41.300$\pm$7.439~$\mathbf{+}$
& 27.658$\pm$4.325~$\mathbf{+}$
& 1529.336$\pm$435.616~$\mathbf{=}$ \\

PSO\_init
& 32.909$\pm$4.993~$\mathbf{+}$
& 27.497$\pm$4.044~$\mathbf{+}$
& 624.382$\pm$443.653~$\mathbf{+}$
& 71.894$\pm$12.446~$\mathbf{+}$
& 59.618$\pm$7.703~$\mathbf{+}$
& 45.985$\pm$6.517~$\mathbf{+}$
& 627.100$\pm$454.390~$\mathbf{+}$
& 1518.345$\pm$0.387~$\mathbf{=}$
& 30.836$\pm$4.892~$\mathbf{+}$
& 442.494$\pm$409.106~$\mathbf{+}$
& 37.367$\pm$6.830~$\mathbf{+}$
& 1833.227$\pm$0.583~$\mathbf{=}$ \\
\hline

random
& 21.109$\pm$3.286~$\mathbf{+}$
& 20.212$\pm$3.211~$\mathbf{+}$
& 39.694$\pm$7.187~$\mathbf{+}$
& 30.376$\pm$5.193~$\mathbf{+}$
& 25.079$\pm$4.131~$\mathbf{+}$
& 26.303$\pm$4.124~$\mathbf{+}$
& 34.258$\pm$5.664~$\mathbf{+}$
& 187.167$\pm$96.433~$\mathbf{+}$
& 20.588$\pm$3.387~$\mathbf{+}$
& 22.179$\pm$4.218~$\mathbf{+}$
& 20.830$\pm$3.344~$\mathbf{+}$
& 40.224$\pm$38.369~$\mathbf{+}$ \\

NP
& 20.000$\pm$3.162~$\mathbf{+}$
& 20.455$\pm$3.602~$\mathbf{+}$
& 41.818$\pm$10.495~$\mathbf{+}$
& 23.545$\pm$4.031~$\mathbf{+}$
& 26.000$\pm$3.162~$\mathbf{+}$
& 27.636$\pm$4.074~$\mathbf{+}$
& 60.545$\pm$18.218~$\mathbf{+}$
& 1513.000$\pm$0.000~$\mathbf{=}$
& 22.182$\pm$4.108~$\mathbf{+}$
& 81.182$\pm$44.995~$\mathbf{+}$
& 26.636$\pm$6.786~$\mathbf{+}$
& 1831.000$\pm$0.000~$\mathbf{=}$ \\

PageRank
& 22.000$\pm$3.162~$\mathbf{+}$
& 22.091$\pm$4.055~$\mathbf{+}$
& 31.455$\pm$7.809~$\mathbf{+}$
& 37.091$\pm$5.915~$\mathbf{+}$
& 25.545$\pm$3.602~$\mathbf{+}$
& 27.273$\pm$5.327~$\mathbf{+}$
& 29.182$\pm$3.433~$\mathbf{+}$
& 63.273$\pm$3.670~$\mathbf{+}$
& 20.000$\pm$3.162~$\mathbf{+}$
& 22.000$\pm$3.162~$\mathbf{+}$
& 22.273$\pm$3.519~$\mathbf{+}$
& 24.000$\pm$3.162~$\mathbf{+}$ \\
\hline

HHD
& 23.091$\pm$3.315~$\mathbf{+}$
& 20.636$\pm$3.574~$\mathbf{+}$
& 93.818$\pm$26.842~$\mathbf{+}$
& 43.455$\pm$7.774~$\mathbf{+}$
& 47.000$\pm$8.146~$\mathbf{+}$
& 36.545$\pm$9.680~$\mathbf{+}$
& 108.455$\pm$18.593~$\mathbf{+}$
& 1513.000$\pm$0.000~$\mathbf{=}$
& 23.727$\pm$4.433~$\mathbf{+}$
& 47.636$\pm$9.306~$\mathbf{+}$
& 27.364$\pm$8.127~$\mathbf{+}$
& 1831.000$\pm$0.000~$\mathbf{=}$ \\

HCI1
& 28.364$\pm$5.262~$\mathbf{+}$
& 22.000$\pm$3.162~$\mathbf{+}$
& 116.182$\pm$25.654~$\mathbf{+}$
& 71.909$\pm$13.918~$\mathbf{+}$
& 46.636$\pm$6.637~$\mathbf{+}$
& 48.818$\pm$8.255~$\mathbf{+}$
& 116.455$\pm$17.053~$\mathbf{+}$
& 1513.000$\pm$0.000~$\mathbf{=}$
& 20.000$\pm$3.162~$\mathbf{+}$
& 34.182$\pm$4.877~$\mathbf{+}$
& 21.818$\pm$3.664~$\mathbf{+}$
& 1339.091$\pm$803.289~$\mathbf{=}$ \\

HCI2\_ICM
& 41.000$\pm$6.325~$\mathbf{+}$
& 27.000$\pm$3.162~$\mathbf{+}$
& 934.000$\pm$862.763~$\mathbf{+}$
& 91.636$\pm$13.439~$\mathbf{+}$
& 63.727$\pm$10.028~$\mathbf{+}$
& 60.545$\pm$8.887~$\mathbf{+}$
& 135.545$\pm$16.919~$\mathbf{+}$
& 1513.000$\pm$0.000~$\mathbf{=}$
& 20.000$\pm$3.162~$\mathbf{+}$
& 34.182$\pm$4.877~$\mathbf{+}$
& 21.818$\pm$3.664~$\mathbf{+}$
& 1339.091$\pm$803.289~$\mathbf{=}$ \\
\hline

\textbf{HDPSO}
& \textbf{58.115$\pm$10.324}
& \textbf{44.764$\pm$8.616}
& \textbf{1829.585$\pm$123.977}
& \textbf{855.536$\pm$768.068}
& \textbf{77.882$\pm$12.244}
& \textbf{63.955$\pm$9.329}
& \textbf{1590.000$\pm$0.000}
& \textbf{1513.000$\pm$0.000}
& \textbf{36.921$\pm$6.632}
& \textbf{1274.703$\pm$561.610}
& \textbf{53.942$\pm$11.354}
& \textbf{1831.000$\pm$0.000} \\
\hline
\end{tabular}%
}
\end{table*}

\begin{table}[htbp]
\centering
\caption{Results of HDPSO and Baseline Methods on Real-World Hypergraphs}
\label{tab:real}
\renewcommand{\arraystretch}{1.2}
\setlength{\tabcolsep}{10pt}
\resizebox{\linewidth}{!}{%
\begin{tabular}{lcccc}
\toprule
\textbf{Algorithm} & \textbf{Algebra} & \textbf{IJO1366} & \textbf{Music-Blue} & \textbf{Geometry} \\
\midrule
GA & 30.87$\pm$5.19\rlap{~$\mathbf{+}$} & 119.93$\pm$9.12\rlap{~$\mathbf{+}$} & 164.73$\pm$24.23\rlap{~$\mathbf{+}$} & 497.87$\pm$182.81\rlap{~$\mathbf{+}$} \\
PSO & 20.00$\pm$0.00\rlap{~$\mathbf{+}$} & 20.20$\pm$0.66\rlap{~$\mathbf{+}$} & 174.47$\pm$4.58\rlap{~$\mathbf{+}$} & 396.87$\pm$246.13\rlap{~$\mathbf{+}$} \\
PSO\_init & 34.00$\pm$1.78\rlap{~$\mathbf{+}$} & 87.70$\pm$4.38\rlap{~$\mathbf{+}$} & 168.43$\pm$1.79\rlap{~$\mathbf{+}$} & 568.43$\pm$0.68\rlap{~$\mathbf{=}$} \\

random & 20.00$\pm$0.00\rlap{~$\mathbf{+}$} & 20.00$\pm$0.00\rlap{~$\mathbf{+}$} & 20.07$\pm$0.37\rlap{~$\mathbf{+}$} & 20.03$\pm$0.18\rlap{~$\mathbf{=}$} \\
NP & 21.00$\pm$0.00\rlap{~$\mathbf{+}$} & 87.00$\pm$0.00\rlap{~$\mathbf{+}$} & 20.00$\pm$0.00\rlap{~$\mathbf{+}$} & 567.00$\pm$0.00\rlap{~$\mathbf{=}$} \\
PageRank & 21.00$\pm$0.00\rlap{~$\mathbf{+}$} & 43.00$\pm$0.00\rlap{~$\mathbf{+}$} & 22.00$\pm$0.00\rlap{~$\mathbf{+}$} & 567.00$\pm$0.00\rlap{~$\mathbf{=}$} \\

HHD & 20.00$\pm$0.00\rlap{~$\mathbf{+}$} & 75.00$\pm$0.00\rlap{~$\mathbf{+}$} & 20.00$\pm$0.00\rlap{~$\mathbf{+}$} & 567.00$\pm$0.00\rlap{~$\mathbf{=}$} \\
HCI1 & 20.00$\pm$0.00\rlap{~$\mathbf{+}$} & 34.00$\pm$0.00\rlap{~$\mathbf{+}$} & 168.00$\pm$0.00\rlap{~$\mathbf{+}$} & 567.00$\pm$0.00\rlap{~$\mathbf{=}$} \\
HCI2\_ICM & 20.00$\pm$0.00\rlap{~$\mathbf{+}$} & 33.00$\pm$0.00\rlap{~$\mathbf{+}$} & 168.00$\pm$0.00\rlap{~$\mathbf{+}$} & 567.00$\pm$0.00\rlap{~$\mathbf{=}$} \\
\midrule
\textbf{HDPSO} & \textbf{100.47$\pm$43.36} & \textbf{143.03$\pm$5.85} & \textbf{834.90$\pm$159.57} & \textbf{567.00$\pm$0.00} \\
\bottomrule
\end{tabular}%
}
\end{table}

As shown in Table~\ref{tab:synthetic}, HDPSO significantly outperforms all the baseline methods on the synthetic datasets, with performance comparable only to the standard GA on the UF hypergraph. This is primarily due to the relatively simple structure of the UF hypergraph. This simplicity results in a smaller search space, enabling the GA to find competitive solutions. On the real-world hypergraphs, according to Table~\ref{tab:real}, HDPSO also consistently outperforms the baseline methods on most datasets. The sole exception is the Geometry dataset, where the performance gap is narrower. These findings further confirm the effectiveness and robustness of the proposed method on both the synthetic and real-world hypergraphs.

\begin{table}[htbp]
\centering
\caption{The statistical results of significance tests on synthetic and real-world hypergraphs}
\label{tab:significance_test}
\renewcommand{\arraystretch}{1.2}
\setlength{\tabcolsep}{8pt}
\resizebox{\linewidth}{!}{%
\begin{tabular}{lccccc}
\toprule
\textbf{Methods} & \textbf{ER (+/=/--)} & \textbf{SF (+/=/--)} & \textbf{UF (+/=/--)} & \textbf{Real-world (+/=/--)} & \textbf{Total (+/=/--)} \\
\midrule
GA            & 4/0/0  & 2/2/0  & 2/1/1  & 4/0/0  & 12/3/1 \\
PSO           & 4/0/0  & 3/1/0  & 3/1/0  & 4/0/0  & 14/2/0 \\
PSO\_init     & 4/0/0  & 3/1/0  & 3/1/0  & 3/1/0  & 13/3/0 \\
random        & 4/0/0  & 4/0/0  & 4/0/0  & 4/0/0  & 16/0/0 \\
NP            & 4/0/0  & 3/1/0  & 3/1/0  & 3/1/0  & 13/3/0 \\
PageRank      & 4/0/0  & 4/0/0  & 4/0/0  & 3/1/0  & 15/1/0 \\
HHD           & 4/0/0  & 3/0/1  & 3/1/0  & 3/1/0  & 13/2/1 \\
HCI1          & 4/0/0  & 3/0/1  & 4/0/0  & 3/1/0  & 14/1/1 \\
HCI2\_ICM     & 4/0/0  & 3/0/1  & 4/0/0  & 3/1/0  & 14/1/1 \\
\midrule
\textbf{Total} & \textbf{36/0/0} & \textbf{28/5/3} & \textbf{30/5/1} & \textbf{30/6/0} & \textbf{124/16/4} \\
\bottomrule
\end{tabular}%
}
\end{table}

\begin{table*}[!t]
\centering
\caption{Summary of the Results on Synthetic and Real-world Hypergraphs}
\label{tab:overall_summary}
\renewcommand{\arraystretch}{1.5}
\setlength{\tabcolsep}{7pt}
\resizebox{\textwidth}{!}{%
\begin{tabular}{|l|cccc|cccc|}
\hline
\multirow{2}{*}{\textbf{Algorithm}} 
& \multicolumn{4}{c|}{\textbf{Synthetic Hypergraphs}} 
& \multicolumn{4}{c|}{\textbf{Real-world Hypergraphs}} \\
\cline{2-9}
& \textbf{Avg. Mean Value} & \textbf{Avg. Best Value} & \textbf{Avg. Mean Rank} & \textbf{Avg. Best Rank}
& \textbf{Avg. Mean Value} & \textbf{Avg. Best Value} & \textbf{Avg. Mean Rank} & \textbf{Avg. Best Rank} \\
\hline
GA        & 476.783 & 655.272 & 2.083 & 2.250 & 159.667 & 215.400 & 3.500 & 3.750 \\
PSO       & 294.515 & 333.122 & 5.333 & 5.083 & 83.483  & 139.733 & 7.000 & 7.500 \\
PSO\_init & 445.971 & 652.956 & 3.167 & 3.167 & 199.783 & 215.650 & 2.000 & 2.250 \\
random    & 40.668  & 67.156  & 9.167 & 9.000 & 15.252  & 21.142  & 8.250 & 8.250 \\
NP        & 307.833 & 321.417 & 6.917 & 6.917 & 160.114 & 173.750 & 4.750 & 4.750 \\
PageRank  & 28.848  & 35.167  & 8.750 & 8.750 & 156.045 & 163.250 & 5.750 & 5.000 \\
HHD       & 317.977 & 330.917 & 5.667 & 5.583 & 157.341 & 170.500 & 5.500 & 5.500 \\
HCI1      & 281.538 & 336.917 & 5.917 & 5.500 & 98.250  & 197.250 & 6.000 & 4.750 \\
HCI2\_ICM & 356.795 & 485.417 & 4.667 & 4.250 & 98.205  & 197.000 & 6.250 & 5.000 \\
\hline
\textbf{HDPSO} & \textbf{769.117} & \textbf{901.869} & \textbf{1.500} & \textbf{1.500}
               & \textbf{314.265} & \textbf{411.350} & \textbf{1.250} & \textbf{1.250} \\
\hline
\end{tabular}%
}
\end{table*}


TABLE~\ref{tab:significance_test} summarizes the statistical significance test results. Overall, HDPSO shows a clear superiority over the baseline methods in most cases. Across the total 144 pairwise comparisons, HDPSO achieves significantly better or statistically similar performance in 140 cases. In particular, on the ER hypergraphs, HDPSO consistently outperforms the compared methods in all 36 comparisons. On the SF and UF hypergraphs, HDPSO remains highly competitive, achieving significantly better or statistically similar results in 33 out of 36 cases and 35 out of 36 cases, respectively. On real-world hypergraphs, HDPSO also achieves strong performance across all 36 comparisons, outperforming the baseline methods significantly in 30 cases and achieving statistically similar results in the other 6 cases.  From a method-specific perspective, HDPSO achieves significantly better performance than random in all 16 cases, and shows no inferior result against PSO, PSO\_init, PageRank or NP. Even when compared with relatively stronger baseline methods such as HHD, HCI1, and HCI2\_ICM, HDPSO still maintains a clear superiority in most cases. These results verify the robustness and stability of the proposed method across different synthetic and real-world hypergraphs.

TABLE~\ref{tab:overall_summary} further reports the overall performance summary on the synthetic hypergraphs. HDPSO achieves the highest Avg. Mean Value and Avg. Best Value, reaching 769.117 and 901.869, respectively. These results are substantially better than those of all the baseline methods. Besides, HDPSO obtains the lowest Avg. Mean Rank and Avg. Best Rank, both equal to 1.500, indicating that it consistently ranks first among all the compared methods. On the real-world hypergraphs, HDPSO again achieves the best overall performance, with the highest Avg. Mean Value and Avg. Best Value of 314.265 and 411.350, respectively. Moreover, its Avg. Mean Rank and Avg. Best Rank are both 1.250, which is also the best among all the compared methods. These results further confirm that HDPSO maintains the best overall optimization capability across different types of hypergraphs.

The superior performance of HDPSO mainly stems from its hypergraph-oriented discrete search mechanism, which enables more effective exploration of the seed selection space. Moreover, the proposed initialization strategy provides high-quality initial particles, while the local search strategy further refines promising candidate solutions during the evolutionary process. Therefore, HDPSO can better balance exploration and exploitation, thereby achieving stronger and more stable performance on different types of hypergraphs. Although HDPSO performs worse than the competing methods in a few cases, mainly on several SF and UF instances, the overall results still consistently demonstrate its effectiveness and competitiveness.



\subsection{Parameter Sensitivity Analysis}

\begin{figure}[htbp]
    \centering
    \subfloat[\label{fig:sensitive_a}]{
        \includegraphics[width=0.47\columnwidth]{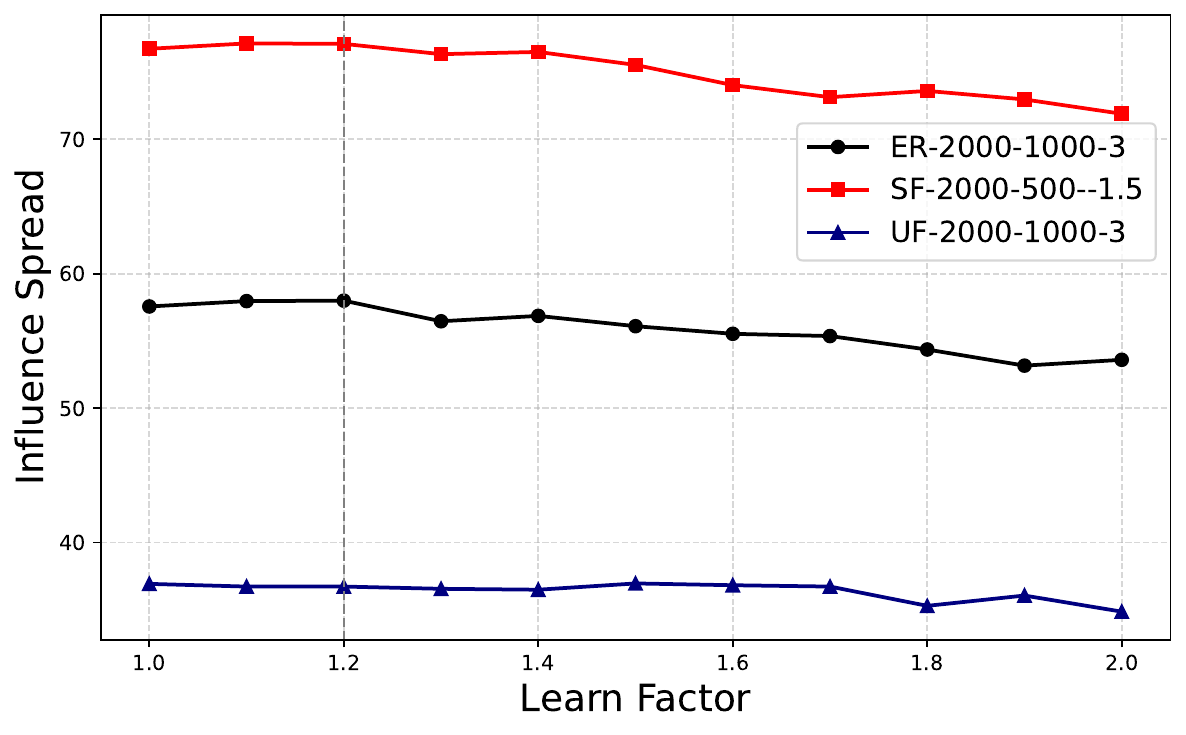}
    }\hfill
    \subfloat[\label{fig:sensitive_b}]{
        \includegraphics[width=0.47\columnwidth]{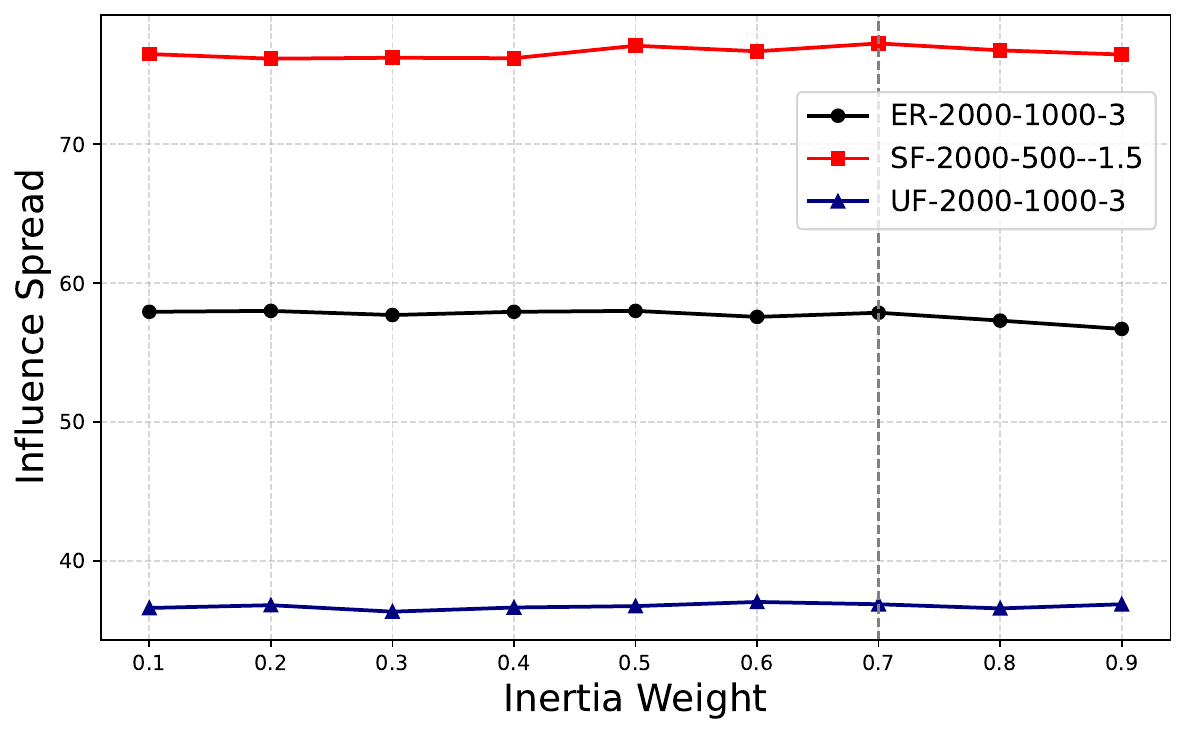}
    }
    \caption{Parameter sensitivity of the HDPSO method on ER, SF, and K-UF hypergraphs. (a) Learning factors \(c_1\) and \(c_2\). (b) Inertia weight \(w\).}
    \label{fig:sensitive}
\end{figure}

In HDPSO, the learning factors \(c_1\) and \(c_2\), along with the inertia weight \(w\), are critical parameters that may significantly affect the algorithm’s performance. We conduct further experiments on the ER, SF, and K-UF hypergraphs to evaluate the sensitivity of HDPSO to these parameters. The corresponding results are presented in Fig.~\ref{fig:sensitive}. In Fig.~\ref{fig:sensitive_a}, the inertia weight is fixed at \(w = 0.5\), while \(c_1\) and \(c_2\) vary from 1 to 2 in steps of 0.1. In contrast, in Fig.~\ref{fig:sensitive_b}, we fix \(c_1 = c_2 = 0.5\), and vary \(w\) from 0.1 to 0.9 in steps of 0.1.

The experimental results reveal that the performance of HDPSO remains relatively stable under different parameter configurations. Specifically, increasing \(w\) enhances the influence of the previous velocity, thereby preserving the particle's momentum and promoting convergence. However, excessively large \(w\) values may result in over-exploration, potentially hindering convergence to the optimal solution. In our experiments, the best performance is achieved when \(w = 0.7\), which strikes a favorable balance between exploration and exploitation.
The learning factors \(c_1\) and \(c_2\) are vital for guiding the search process. Our empirical observations indicate that a configuration of \(c_1 = c_2 = 1.2\) provides a stable search dynamic and consistently good performance across different hypergraph structures. 

\section{Conclusions}


This paper aims to identify a subset of seed nodes with the highest potential to propagate influence throughout a hypergraph-modeled network. The goal has been successfully achieved by proposing a novel method named HDPSO. In HDPSO, the threshold model is employed to simulate the cascading propagation process, and PSO is introduced to identify the most influential nodes in the hypergraph. To improve the quality of initial solutions, a degree-based initialization strategy is adopted, enabling the method to accelerate the convergence. To efficiently evaluate the fitness of particles (i.e., candidate seed sets), we define a two-layer local influence approximation to estimate the spread of each particle. In addition, velocity and position update rules are improved to effectively guide the particles toward higher-quality solutions over successive generations. 

Experimental results on both the synthetic and real-world hypergraphs demonstrate that HDPSO consistently outperforms the baseline methods. The performance advantage of HDPSO becomes more pronounced as the number of selected seed nodes increases. However, due to the intrinsic stochasticity and sensitivity of the threshold model, the propagation processes exhibit a degree of instability. Future research would focus on exploring the dynamics of cascading propagation under various threshold settings. 


\bibliographystyle{IEEEtran}
\bibliography{Bibliography}

\begin{IEEEbiography}[{\includegraphics[width=1in,height=1in]{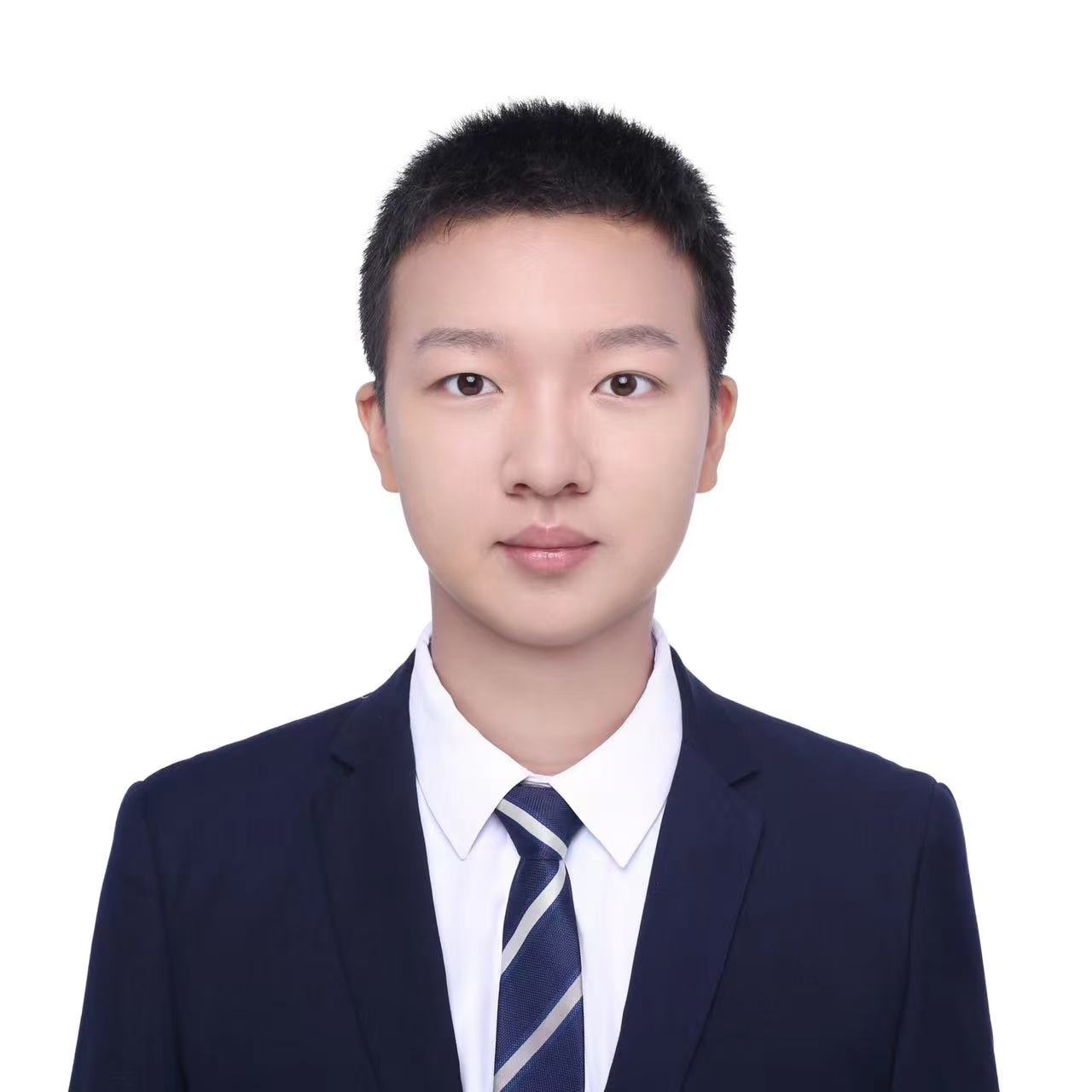}}]{Qianshi Wang} is pursuing a Bachelor's degree at International School of Information Science \& Engineering, Dalian University of Technology. His research interests include hypergraphs, influence maximization, and evolutionary computation.
\end{IEEEbiography}

\begin{IEEEbiography}[{\includegraphics[width=1in,height=1.4in]{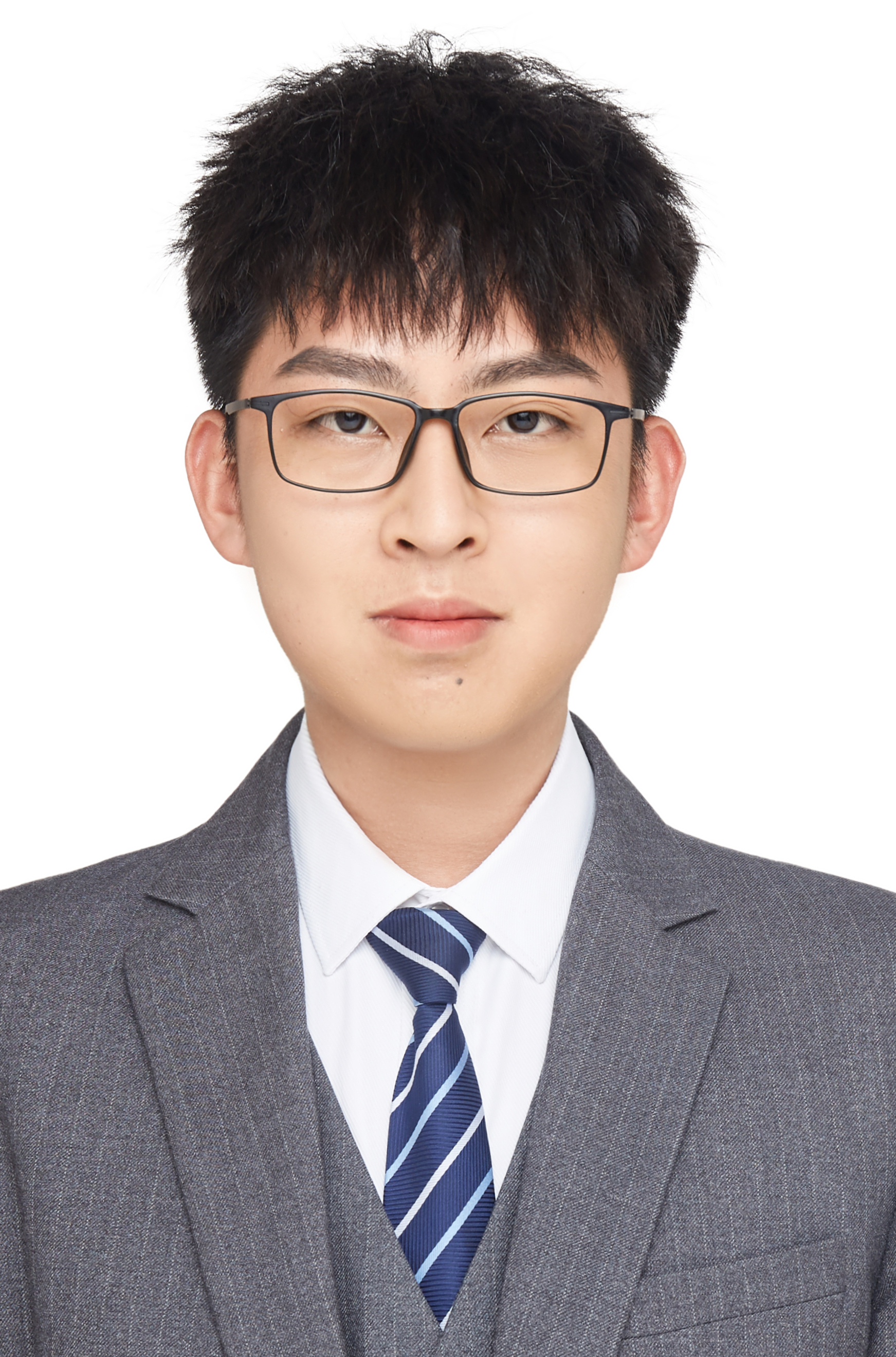}}]{Xilong Qu} received his Bachelor's degree from the School of Mathematics and Statistics at Shandong University of Technology in 2022. He is pursuing a PhD at the School of Computer Science and Technology at Dalian University of Technology. His research interests include complex systems, influence maximization, and evolutionary computation. He has been serving as a reviewer for international journals, including IEEE Transactions on Neural Networks and Learning Systems, Information Processing \& Management, and Applied Soft Computing.  
\end{IEEEbiography}


\begin{IEEEbiography}[{\includegraphics[width=0.9in,height=1.2in]{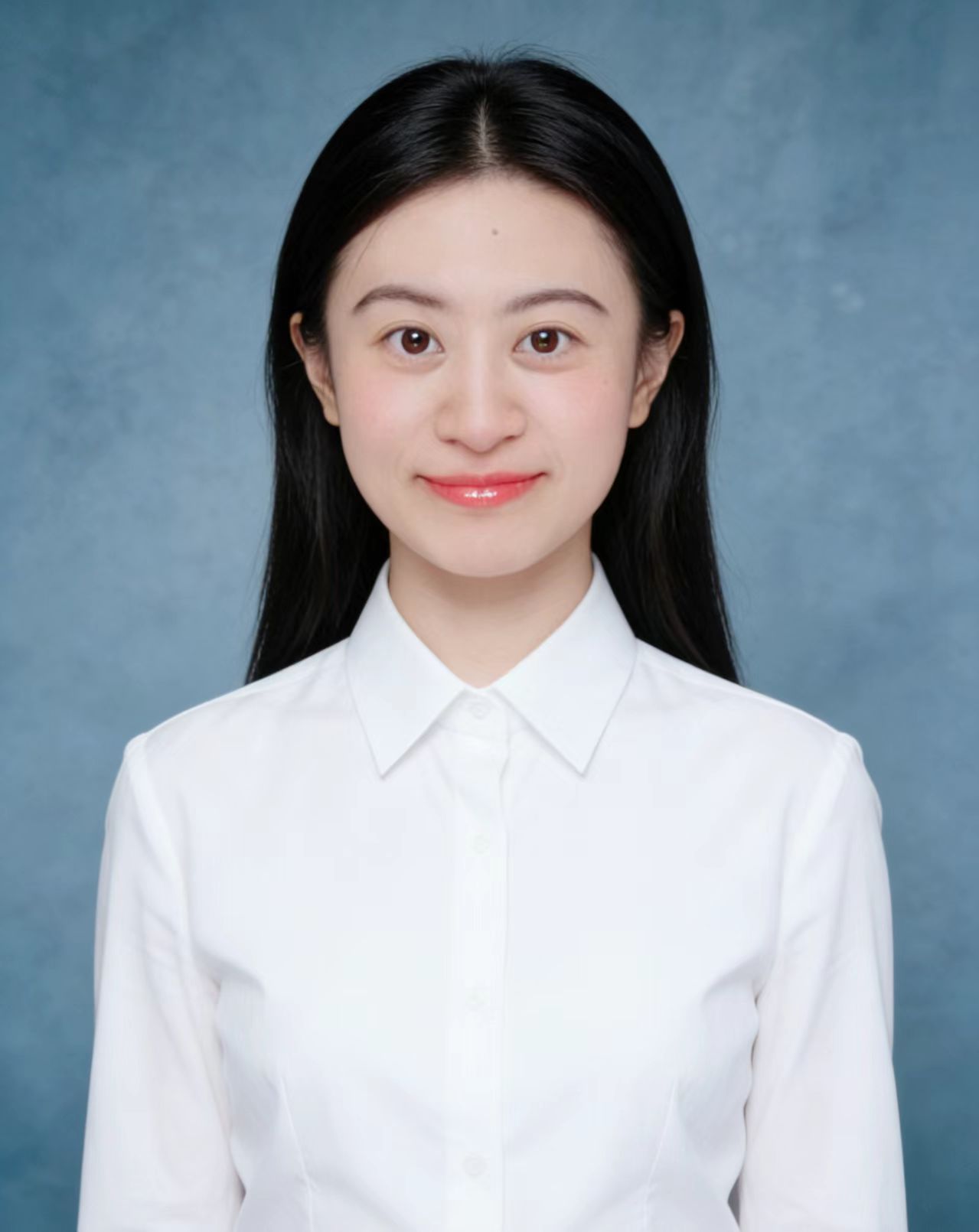}}]{Wenbin Pei} received her Ph.D. degree at Victoria University of Wellington, New Zealand, in 2021. She is currently an assistant professor at Dalian University of Technology. Her research interests include evolutionary computation and machine learning. She has published over 40 papers in fully refereed international journals and conferences, including IEEE Transactions on Evolutionary Computation and Evolutionary Computation journal (MIT Press). She was a program committee member for 7 international conferences and a co-chair of special sessions in 5 international conferences. She has been serving as a reviewer for international journals, including IEEE Transactions on Evolutionary Computation, IEEE Transactions on Cybernetics, Evolutionary Computation journal (MIT Press), IEEE Transactions on Emerging Topics in Computational Intelligence, Applied Soft Computing, and Artificial Intelligence in Medicine. %
\end{IEEEbiography}

\begin{IEEEbiography}[{\includegraphics[width=0.9in,height=1.2in]{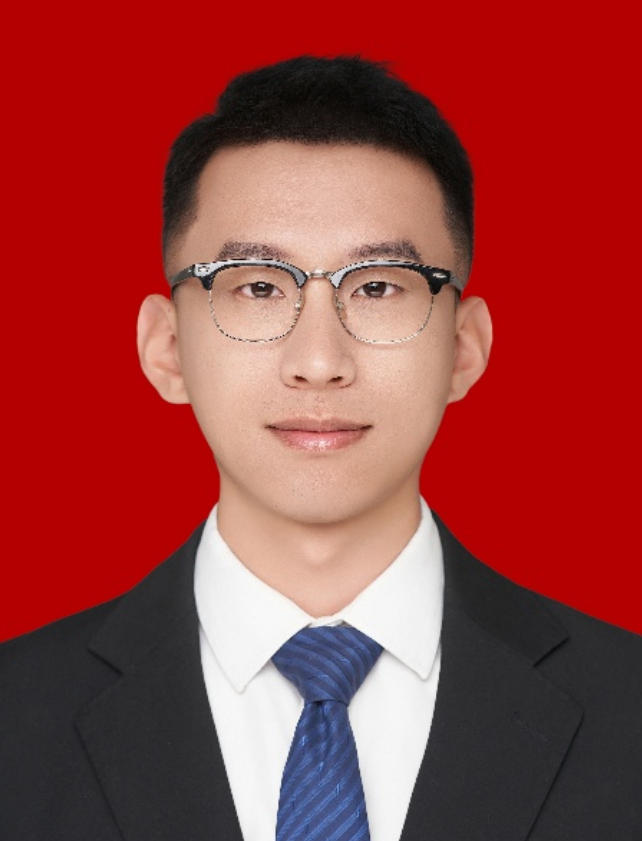}}]{Nan Li} received the B.Sc.
degree in software engineering from North
University of China, Taiyuan, China, in
2020, and the Ph.D. degree in software
engineering from Northeastern University,
Shenyang, China, in 2025. He is currently
a lecturer with the School of Artificial Intelligence, Shanxi University, Taiyuan, China.
His research interests include computational intelligence and machine learning. Dr. Li serves as a Reviewer for international
conferences, including ICML and IJCAI, and international journals, including IEEE TEVC and IEEE TNNLS.
\end{IEEEbiography}

\begin{IEEEbiography}[{\vspace{-4mm} \includegraphics[width=1in,height=1.4in]{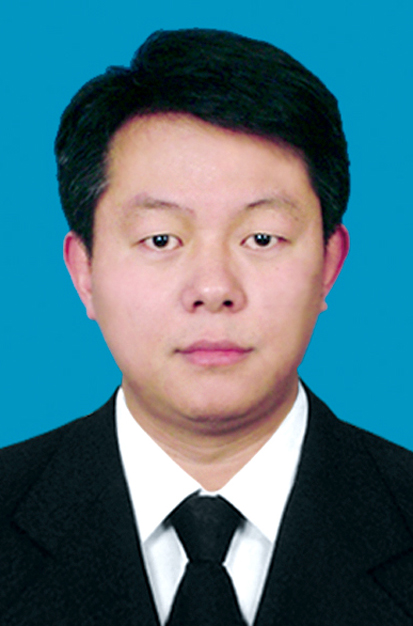}}]{Qiang Zhang} received the Ph.D. degree in Circuits and Systems from Xidian University, Xi'an, in 2002. He is currently a professor and the dean of the School of Computer Science and Technology at Dalian University of Technology. His research interests include bio-inspired computing and the related applications. Prof. Zhang has published more than 500 papers in fully refereed international journals and conferences. He was awarded the National Science Fund for Distinguished Young Scholars in 2014, and was also selected as one of the state department special allowance experts. He has been serving as an editorial board member for 7 international journals and chairs of special issues in journals, such as Neurocomputing and International Journal of Computer Applications in Technology.
\end{IEEEbiography}
\end{document}